\documentclass[preprint,nofootinbib,nobibnotes,groupaddress,preprintnumbers,aps,prd]{revtex4}
\usepackage{bm,epsfig,mathrsfs,amsmath,amssymb,graphicx}
\usepackage{epsfig,amsmath,graphicx,amssymb}
\usepackage{graphicx}
\usepackage{dcolumn}
\usepackage{bm}
\usepackage[toc,title,titletoc,header]{appendix}
\def\dbar{{\mathchar'26\mkern-12mu d}}

\begin{document}

\title{Coefficient of performance under  maximum $\chi$ criterion in a two-level atomic system as a refrigerator}
\author{Yuanyuan$^{1}$}
\author{Rui Wang$^{1}$}
\author{Jizhou He$^{1}$}
\author {Yongli Ma$^2$}
\author{Jianhui Wang$^{1,2}$} \email{wangjianhui@ncu.edu.cn}
\affiliation{ $^1\,$ Department of Physics, Nanchang University,
Nanchang 330031, China \\ $^2\,$ State Key Laboratory of Surface
Physics and Department of Physics, Fudan University, Shanghai
200433, China}

\begin{abstract}
A two-level atomic system as a working substance is  used to set up
a refrigerator consisting of two quantum  adiabatic and two
isochoric processes (two constant-frequency processes $\omega_a$ and
$\omega_b$ with $\omega_a<\omega_b$), during which the two-level
system is in contact with two heat reservoirs at temperatures $T_h$
and $T_c (<T_h)$. Considering finite-time operation of two isochoric
processes, we derive analytical expressions for cooling rate $R$ and
coefficient of performance  (COP) $\varepsilon$. The COP at maximum
$\chi(= \varepsilon R)$ figure of merit is numerically determined,
and it is proved to be in nice agreement with the so-called Curzon
and Ahlborn  COP $\varepsilon_{CA}=\sqrt{1+\varepsilon_C}-1$, where
$\varepsilon_C=T_c/(T_h-T_c)$ is the Carnot COP. In the
high-temperature limit,  the COP at maximum $\chi$ figure of merit,
$\varepsilon^*$, can be expressed analytically by
$\varepsilon^*=\varepsilon_{+}\equiv({\sqrt{9+8\varepsilon_C}-3})/2$,
which was derived previously as  the upper bound of optimal COP for
the low-dissipation or minimally nonlinear irreversible
refrigerators. Within context of irreversible thermodynamics, we
prove that the value of $\varepsilon_{+}$ is also the upper bound of
COP at maximum $\chi$ figure of merit when we regard our model as  a
linear irreversible refrigerator.

 PACS number(s): 05.70.Ln

\end{abstract}

\maketitle
\date{\today}

\section {introduction}
A heat device is a heat engine that convert thermal energy into
mechanical work, or a refrigerator (heat pump) that is basically a
heat engine running backwards. For an endoreversible heat engine
working between a hot and a cold reservoir at constant temperatures
$T_h$ and $T_c (<T_h)$,  Curzon and Ahlborn (CA) \cite{Cur75} found
the efficiency of at maximum power to be
$\eta_{CA}=1-\sqrt{T_c/T_h}=1-\sqrt{1-\eta_C}$ with
$\eta_C=1-T_c/T_h$  the Carnot efficiency.  The model of  such a
heat engine presented by Curzon and Ahlborn gave rise to the birth
and intensive studies of finite-time thermodynamics \cite{And77,
Ber77, Gut78, Rub80, Vos85, Chen89, Gev92, Chen94, Bej96, Bro05,
Rez06, Sheng14, Wu04}, a branch of thermodynamics focusing on the
optimization on the energy converter that consists of some
finite-time thermodynamic processes.
 The universality and bounds \cite{Bro05, Esp10, Espo10, Wang12, All13, Jh1203, Jh1205, All08, Izu10, Guo13} of
the efficiency at maximum power  have been discussed  in a large
number of studies of heat engines within the context of finite-time
thermodynamics.

Unlike in analysis of a heat engine where the power output is always
an objective function to determine the optimized efficiency, there
are various optimization criteria \cite{Vel97, Vela97, Yan90, Tom13,
Tom12, Izu13, Hu13, Tu12} in analysis of optimization of a
refrigerator working between two heat reservoirs with constant
temperatures $T_h$ and $T_c$. One of these criteria for a
refrigerator, which was first proposed by Yan and Chen \cite{Yan90},
is taking the target function $\chi=\varepsilon Q_c/t_{cycle}$,
where $Q_c$ is heat absorbed from the cold reservoir, $t_{cycle}$
denotes the cycle time, and $\varepsilon = Q_c/W$ with $W$ being the
work input per cycle is the coefficient of performance (COP) for
refrigerators, This $\chi-$optimization criterion for refrigerators
is always adopted and found to be exactly the counterpart
\cite{Tu12, Tom12, Tom13} for the optimization of power output for
heat engines. For a low dissipation
  \cite{Tu12, Hu13} or a minimally nonlinear irreversible \cite{Izu13}
refrigerator, the lower and upper bounds of the COP at maximum
$\chi$ figure of merit ($\varepsilon^*$) have been found to be:  $0
\le \varepsilon ^*\le \left( {\sqrt {9 + 8\varepsilon _C } - 3}
\right)/2$, with $\varepsilon_C=T_c/(T_h-T_c)$ being so-called
Carnot COP.

The research into heat engines or refrigerators has been extend from
classical to quantum systems \cite{Ben00, Sco59, Rez06, Gev92,
Quan09, Rui13, Rut09, Rui12, Fel00,  Huang13, Cor13, Low13, Aba12,
Lev12, All10, Cle12, Jar12, Jar13,Scu02, 8504} over 50 years. This
is motivated by exploring the emergence of basic thermodynamic
description at the quantum mechanical level, and also by the
potential technological applications of these devices \cite{Low13,
Bli12, Aba12, Cle12, Rut09}. In particular, demands for smaller heat
devices have been rapidly rising because of miniaturization in
experiment \cite{Low13, Bli12} and understanding of quantum
thermodynamics \cite{Jar12, Jar13}. The ongoing reduction in system
size is approaching the ultimate limit, scaling downing these heat
devices to a single particle system, in which quantum properties
become significant and have thus to be fully considered.

As quantum versions of classical thermodynamic cycles, quantum
thermodynamic cycles have also a set of different cyclic heat device
models \cite{8504} working between heat reservoirs.  The quantum
Otto cycle, which is a typical one of these models and  the quantum
analog of the classical Otto cycle used widely in practical heat
devices, has been proposed and discussed in a series of papers
\cite{Gev92, Scu02, Rez06, Rui13, 8504, Aba12}. The present paper
employs a two-level atomic system as a working substance to set up a
refrigerator model, which consists of two isochoric and two
adiabatic processes and is thus a quantum version of the Otto
refrigeration cycle. Based on master equations of stochastic
processes, we derive expressions for the cooling rate and power
input, which are functions of the time allocation on the two
isochores. The objective function $\chi$ is then numerically
optimized to determine the optimal COP $\varepsilon^*$, which is
also analytically expressed as a function of Carnot COP
$\varepsilon_C$ in the high-temperature limit. Finally, we analyze
the COP at maximum $\chi$ figure of merit by taking our model as a
linear irreversible refrigerator satisfying the tight-coupling
condition.

\section{A model of quantum otto refrigeration cycle} \label{isoen}
\subsection{Dynamics of occupation probabilities}
 In the refrigerator model  the working substance is a two-level
energy system, with ground state $g$ and excited state $e$
characterized by the energy spectrum $\varepsilon_g=\omega$ and $
\varepsilon_e=2\omega$ ($\hbar\equiv1$) and by the energy gap
$\Delta\varepsilon=\varepsilon_e-\varepsilon_g=\omega$. Let $p_g$
and $p_e$ denote
the occupation probabilities of the two states $e$ and $g$, and these probabilities must satisfy the constraint $p_e+p_g=1$. 

When  a two-level energy system is coupled to a heat reservoir at
constant temperature $T=1/\beta$ $(k_B\equiv1)$, the dynamics of the
occupation probabilities at the ground and excited states, $p_g$ and
$p_e$, can be determined according to the following master equation
\cite{Fel00, Rui13}:
\begin{equation}
\dot{\textbf{p}}(t)=R\cdot\textbf{p}(t), \label{ptpt}
\end{equation}
 with $\textbf{p}(t)=(p_g,
p_e)^{\mathrm{T}}$(where the superscript $\mathrm{T}$ denotes
transpose). Here the stochastic matrix $R$ describing particle
dynamics is given by
\begin{equation}
 R=\left(
\begin{array}{c c}
-k_\uparrow  & k_\downarrow \\ k_\uparrow  & -k_\downarrow
\end{array}
\right), \label{r}
\end{equation}
where  $k_\downarrow$ and $k_\uparrow$ represent the transition
rates from the excited to the ground level and vice versa, and they
satisfy the requirement of detailed balance \cite{Jar13, Kam07}:
\begin{equation}
\frac{k_\uparrow}{k_\downarrow}=e^{-\beta\Delta \varepsilon}.
\label{frvar}
\end{equation}
Then these transition rates can be parameterized by \cite{Jar13}
\begin{equation}
k_\downarrow =\gamma(1-\sigma), k_\uparrow =\gamma(1+\sigma),
\label{k}
\end{equation}
with $\sigma=\tanh(\beta\Delta\varepsilon/2)$, where $\gamma>0$
denotes a characteristic rate for these transitions and it will be
identified as the heat conductivity in the following [below Eq.
(\ref{dn})].

%

Now we turn to the discussion of the refrigerator model operating in
finite time. The working substance of the quantum Otto refrigerator
is a two-level atomic system with time-dependent energy unit $\omega
(t)$, changing between $\omega_a$ and $\omega_b$.  The two level
system is alternatingly coupled to two heat baths at inverse
temperatures $\beta_c$ and $\beta_h (<\beta_c)$. The Otto cycle
consists of four consecutive steps shown in Fig. \ref{model} and it
is described as follows:
\begin{figure}[tb]
\includegraphics[width=2.8in]{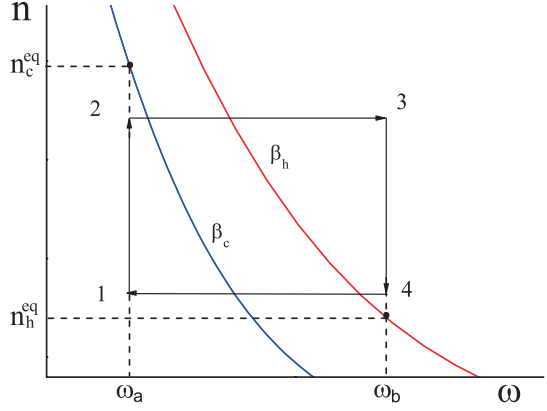}
 \caption{(Color online)  Schematic diagram of a quantum Otto refrigeration cycle in the $(\omega, n)$
 plane. $1\rightarrow 2$ and $3\rightarrow 4$ are two isochoric processes,
while $2\rightarrow 3$ and $4\rightarrow 1$ are two adiabatic
processes.
 ${n}_h^{eq}$ and ${n}_c^{eq}$ are  two ratios when the atomic system
achieves thermal equilibrium with two heat reservoirs at  inverse
temperatures $\beta_h$ and $\beta_c$, respecitively.}\label{model}
\end{figure}

(1) Cold isochore $1\rightarrow 2$. Initially at time $t=0$, the
system becomes coupled to a cold reservoir at inverse temperature
$\beta_c$ and is decoupled from this reservoir until time
$t=\tau_c$, while the frequency is kept constant $\omega_a$. From
Eq. (\ref{ptpt}), we obtain the probabilities $\textbf{p}({t})$ at
any instant of the isochore ($0\leq t\leq\tau_c$) as,
\begin{equation}
\textbf{p}({t})=\exp(R_c t)\textbf{p}({0}) \label{ptc}
\end{equation}
where $R_c=\gamma_c\left(\begin{array}{c c} -(1+\sigma_c)&(1-\sigma_c)\\
(1+\sigma_c) &-(1-\sigma_c)
\end{array}\right)$ with $\sigma_c=\tanh(\beta_c\omega_a/2)$.

(\emph{2}) Adiabatic compression $2\rightarrow 3$. The system is
isolated in time $\tau_{ch}$ while its frequency $\omega$ changes
from $\omega_a$ to $\omega_b$, with a slow speed to satisfy the
quantum adiabatic condition. In the  adiabatic process, the entropy
is kept constant as
\begin{equation}
\textbf{p}({t})=\textbf{p}({\tau_c}), \label{pta}
\end{equation}
with ${\tau_c}\le t\le{\tau_c+\tau_{ch}}$.  

{(\emph{3})} Hot isochore $3\rightarrow 4$. The system is now
coupled to a hot reservoir at inverse temperature $\beta_h$
($<\beta_c$) in time of $\tau_h$ and its energy unit is again kept
constant. As in the process $1\rightarrow2$,  the evolution of the
probabilities $\textbf{p}({t})$ at any instant
($\tau_c+\tau_{ch}\leq t\leq\tau_c+\tau_{ch}+\tau_h$ ) during  the
hot isochore can be determined according to
\begin{equation}
\textbf{p}({t})=\exp(R_h t)\textbf{p}({\tau_c+\tau_{ch}}),
\label{ptch}
\end{equation}
where $R_h=\gamma_h\left(\begin{array}{c c} -(1+\sigma_h)&(1-\sigma_h)\\
(1+\sigma_h) &-(1-\sigma_h)
\end{array}\right)$ with $\sigma_h=\tanh(\beta_h\omega_b/2)$.


{(\emph{4})} Adiabatic expansion $4\rightarrow 1$. The energy unit
$\omega$ is changed very slowly (as in the adiabatic compression) to
its initial value $\omega_a$, while the probabilities $\textbf{p}$
are kept unchanged. Let $\tau_{hc}$ be the time taken for completing
this adiabat. When $\tau_c+\tau_{ch}+\tau_h\leq
t\leq\tau_c+\tau_{ch}+\tau_h+\tau_{hc}$,  we have
\begin{equation}
\textbf{p}({t})=\textbf{p}({\tau_c+\tau_{ch}+\tau_h}). \label{pth}
\end{equation}


After a single cycle, the  entropy of the system as a  state
function changes back to its initial value, and therefore we have
$\textbf{p}(t_{cycle})=\textbf{p}(0)$, where
$t_{cycle}\equiv\tau_c+\tau_{ch}+\tau_h+\tau_{hc}$ is the cycle
time. It follows, using Eqs. (\ref{ptc})- (\ref{pth}), that the
probabilities of the final and initial system states during a cycle
satisfy the relation:
\begin{equation}
{\left(
\begin{array}{c}
p_g(t_{cycle})\\ p_e(t_{cycle})
\end{array}
\right)}=\exp(R_h t)\exp(R_c t){\left(
\begin{array}{c}
p_g(0)\\ p_e(0)
\end{array}
\right)}. \label{pg}
\end{equation}
where $R_c$ and $R_h$ were defined in Eqs. (\ref{ptc}) and
(\ref{ptch}), respectively. Let $\mathcal{M}=\exp(R_h t)\exp(R_c t)$
be the transition matrix for the two-level system proceeding a
cycle. Note that, the initial instant of the system per cycle under
consideration can be assumed to be a periodic steady state
\cite{note1}. Considering Eq. (\ref{pg}), we find
\begin{equation}
{\left(
\begin{array}{c}
p_g(0)\\ p_e(0)
\end{array}
\right)}=\mathcal{M}{\left(
\begin{array}{c}
p_g(0)\\ p_e(0)
\end{array}
\right)}, \label{p0}
\end{equation}
from which we obtain the probabilities $\textbf{p}(0)$ at the
initial instant per cycle as
\begin{equation}
\textbf{p}(0) = {\left(
\begin{array}{c}
p_g(0)\\ p_e(0)
\end{array}
\right)}=\left[ {\begin{array}{c}
{\frac{(1-\sigma_h)e^{2(\gamma_c\tau_c +\gamma_h\tau_h)}+(\sigma_h -\sigma_c)
e^{2\gamma_c\tau _c}+\sigma_c- 1}{2e^{2(\gamma_c\tau_c+\gamma_h\tau_h)} - 2}}\\
{\frac{(\sigma_h+1)e^{2(\gamma_c\tau_c
+\gamma_h\tau_h)}+(\sigma_c-\sigma_h)e^{2\gamma_c\tau _c}-\sigma_c-
1}{2e^{2(\gamma_c\tau_c+\gamma_h\tau_h)} - 2}}
\end{array}} \right]. \label{p01}
\end{equation}
Using Eqs. (\ref{ptc}) and (\ref{p01}), we obtain
\begin{equation}
\textbf{p}(\tau_c) = {\left(
\begin{array}{c}
p_g(\tau_c)\\ p_e(\tau_c)
\end{array}
\right)}=\left[ {\begin{array}{c}
{\frac{(1-\sigma_c)e^{2(\gamma_c\tau_c +\gamma_h\tau_h)}+(\sigma_c -\sigma_h)
e^{2\gamma_h\tau _h}+\sigma_h- 1}{2e^{2(\gamma_c\tau_c+\gamma_h\tau_h)} - 2}}\\
{\frac{(\sigma_c+1)e^{2(\gamma_c\tau_c
+\gamma_h\tau_h)}+(\sigma_h-\sigma_c)e^{2\gamma_h\tau _h}-\sigma_h-
1}{2e^{2(\gamma_c\tau_c+\gamma_h\tau_h)} - 2}}
\end{array}} \right].\label{p0h2}
\end{equation}

\subsection{Cooling load and COP}

The amount of change in energy $d E$ for the system follows from the
first law of thermodynamics: $dE=\dbar Q+\dbar
W$=$\sum_{\nu}\varepsilon_{\nu}dp_{\nu}+\sum_{\nu}p_{\nu}d\varepsilon_{\nu}$,
where $\dbar Q=\sum_{\nu}\varepsilon_{\nu}dp_{\nu}$ and $\dbar
W=\sum_{\nu}p_{\nu}d\varepsilon_{\nu}$ are the heat exchange and
work done, respectively. Accordingly, during an adiabatic process
there is no heat exchange $(\dbar Q=0)$ as the occupation
probabilities $p_{\nu}$ do not change, but work may still be nonzero
(since eigenenergies $\varepsilon_{\nu}$ may change). For
simplicity, the total energy of the two-level system ($E=p_g\omega+2
p_e\omega$) can be written as $E=n\omega$, with $n\equiv p_g+2p_e$.
It follows, using the relation $n=p_g+2p_e$, that
\begin{equation}
dE=\dbar W+\dbar Q=nd\omega+\omega dn, \label{dn}
\end{equation}
where $\dbar Q=\omega dn $ and $\dbar W=nd\omega$. The energy of the
system can change either by particle transition (changing ${n}$) or
by varying the energy gap between two states (changing $\omega$).
The heat current during the hot or cold isochoric process is
determined by $\dbar Q =\omega \frac{d n}{dt}=\frac{d
(p_g+2p_e)}{dt}$ which, together with Eqs. (\ref{ptpt}), (\ref{r})
and (\ref{k}), indicates that $\gamma_{c,h}$ represent the heat
conductivities between the working substance and the cold and hot
reservoirs, respectively.

Since there exist no heat exchanged during the two adiabatic
process, we can merely determine the heat exchanged during the cold
and hot isochoric processes, $Q_c$ and $Q_h$, to obtain the COP,
$\varepsilon={Q_c}/{W}$ with work input $W$. In view of the fact
that no work is done in any isochore, the heat absorbed by the
system from the cold reservoir in the cold isochore $1\rightarrow2$,
which is just the cooling load,  can be directly calculated as
\begin{equation}
Q_c=\int_{0}^{\tau_c}\omega_a \frac{d n}{dt}dt=({n}_2-{n}_1)
\omega_a, \label{qc}
\end{equation}
where $n_2=p_g(\tau_c)+2 p_e(\tau_c)$ and $n_1=p_g(0)+2 p_e(0)$.
 Similarly, we can easily derive the amount of heat released to
the hot reservoir during the hot isochore $3\rightarrow4$ as
\begin{equation}
Q_h=\left|\int_{\tau_c+\tau_{ch}}^{\tau_c+\tau_{ch}+\tau_h}\omega_b
 \frac{d n}{dt}dt\right|=({n}_2-{n}_1) \omega_b, \label{qh}
\end{equation}
where the use of $n_3=n_2$ and $n_4=n_1$ has been made (see Fig.
\ref{model}). Then the COP of the quantum refrigerator becomes
\begin{equation}
\varepsilon=\frac{\omega_a}{\omega_b-\omega_a}. \label{var}
\end{equation}

As shown in Eqs. (\ref{pta}) and (\ref{pth}), the occupation
probabilities $\textbf{p}(t)$ are kept unchanged during an adiabatic
process. It follows, using Eqs. (\ref{p01}) and (\ref{p0h2}) as well
as $T=1/\beta$, that in the adiabatic process there exists the
relation $T\omega^{-1}=$const.  Making a comparison of this adiabat
relation to that for classical ideal gas, $TV^{\gamma-1}=$ const,
with $\gamma$ the adiabatic parameter, we find  that the Otto cycle
COP (\ref{var}) for the two level system is analogous to the COP of
the Otto cycle working with the classical ideal gas:
$\varepsilon^{IG}=\left(\frac{V_a}{V_b-V_a}\right)^{\gamma-1}$,
where $V_a$ and $V_b$ ($>V_a$) are the constant volumes along the
two isochores. It is therefore appropriate to  take the frequency
$\omega$ as the volume variable \cite{notev} and to identify this
model as a quantum version of the classical Otto refrigeration
cycle.

\section{The optimization of quantum Otto refrigeration cycle} \label{gen}

Making use of  Eqs. (\ref{p01}), (\ref{p0h2}), and (\ref{qc}), we
obtain the  relation:
${n}_2-{n}_1=({n}_c^{eq}-{n}_h^{eq})\frac{(e^{2\gamma_c\tau_c}-1)
(e^{2\gamma_h\tau_h}-1)}{e^{2\gamma_c\tau_c+2\gamma_h\tau_h}-1}$,
with ${n}_c^{eq}=\frac{1}2[\tanh(\beta_c\omega_a/2)+1]$ and $
{n}_h^{eq}=\frac{1}2[\tanh(\beta_h\omega_b/2)$+1].  When the two
isochores are quasistatic ($\tau_c\rightarrow\infty$ and
$\tau_h\rightarrow \infty$), the system approaches thermal
equilibrium with the cold (hot) reservoir and thus  $n_2$ ($n_1$)
tends to be maximum ( minimum) value $n_c^{eq}$ ($n_h^{eq}$).  The
same values of $n_c^{eq}$($n_h^{eq}$) can also be obtained in a
different method \cite{Rui13}, which is based on the assumption that
the system is at thermal equilibrium. That is, for the system  at
thermal equilibrium with a heat reservoir at constant temperature
$\beta$, the occupation probabilities $p_g$ and $p_e$ satisfy the
Boltzmann distribution: $p_e = p_g e^{-\beta\omega}$, in which
$p_e+p_g=1$. Then occupation probabilities are given by
$p_g=1-p_e=\frac{1}{e^{-\beta\omega}+1}$, giving rising to
${n}^{eq}=\frac{1}2[\tanh(\beta\omega/2)+1]$.

Casting the factor $2$ into $\gamma_{h,c}$ for the expression of
$(n_2-n_1)$, we arrive at
\begin{equation}
{n}_2-{n}_1=\Delta n^{eq} f(\tau_c,\tau_h), \label{n2eq}
\end{equation}
where we have defined $ f(\tau_c,
\tau_h)\equiv\frac{(e^{\gamma_c\tau_c}-1)(e^{\gamma_h\tau_h}-1)}
{e^{\gamma_c\tau_c+\gamma_h\tau_h}-1}$
and $\Delta n^{eq}\equiv n_c^{eq}-n_h^{eq}.$  The expression for
$f(\tau_c, \tau_h)$ is the same as those  derived from heat engines
or refrigerators within different approaches \cite{Rui13, Rez06,
Fel00}, providing a strong argument in favor of our approach.
Additionally, the difference $\Delta n^{eq}$ for the two-level
system is the same as one obtained from a spin-$1/2$ system
\cite{Fel00}. Physically,   the particle in the two-level system
makes transitions between the upper and lower levels by exchanging
energy with the cold or hot reservoir during the interaction
interval, indicating that the two-level system  is in complete
analogy with the spin-$1/2$ system.
 Considering Eqs.
(\ref{qc}), (\ref{var}) and (\ref{n2eq}), and the cycle time
$t_{cycle}=\tau_{adi}+\tau_c+\tau_h$ with
$\tau_{adi}=\tau_{hc}+\tau_{ch}$,  one can write the cooling rate
$R=Q_c/t_{cycle}$ and the objective function $\chi=\varepsilon
R/t_{cycle}$ as
\begin{equation}
R=\frac{f(\tau_c,\tau_h)\omega_a\Delta{n}^{eq}}{\tau_{adi}+\tau_c+\tau_h},
\label{qcuh}
\end{equation}
and
\begin{equation}
\chi=\frac{f(\tau_c,\tau_h)\omega_a^{2}\Delta
n^{eq}}{(\omega_b-\omega_a) (\tau_{adi}+\tau_c+\tau_h)}.
\label{chih}
\end{equation}
From Eqs. (\ref{n2eq}) and (\ref{qcuh}),   the condition for the
interrelation between the temperatures of heat reservoir and the
frequency values can be derived as
$\beta_c/\beta_h<\omega_b/\omega_a$, which must be satisfied in
order that the refrigerator can do cooling and is  the opposite
inequality of positive work condition of the heat engine (see Eq.
(24) of Ref. \cite{Rui13}). In the heat refrigerator work is done on
the working subsystem and thus no useful work is done, thereby
indicating that Carnot's bound is not violated.

The figure of merit, $\chi$, is a product of two functions:
$G(\beta_c,\omega_a, \beta_h, \omega_b)\equiv \omega_a^{2}\Delta
n^{eq}/{(\omega_b-\omega_a)}$, a function merely depends on the
external parameters $\beta$ and $\omega$, and $F\equiv
f(\tau_c,\tau_h)/(\tau_{adi}+\tau_c+\tau_h)$ which describes the
time allocations on the isochores and adiabats. In the case when the
external constraints of the refrigerator are given, optimizing the
objective function $\chi$ is equivalent to optimizing the
time-dependent function $F(\tau_c,\tau_h)$. Because the
probabilities $\textbf{p}(\tau_c,\tau_h)$, which determine the
difference between $n_2$ and $n_1$ and thus determine the objective
function $\chi$ as well as COP $\varepsilon$, are functions of the
time allocation to the two isochores,   the interaction time
($\tau_c$ or $\tau_h$) taken for either of the two isochores as one
of detailed protocols  independently determine $\chi$ as well as
$\varepsilon$.  Setting ${\partial F}/{\partial {\tau_c}}=0$ and
${\partial F}/{\partial {\tau_h}}=0$, the optimal time allocations
on the cold and hot isochores is obtained,
\begin{equation}
\gamma_c[ \cosh(\gamma_h \tau_h) -1] = \gamma_h [\cosh(\gamma_c
\tau_c) -1], \label{gammac}
\end{equation}
which was derived much earlier in Ref. \cite{Rez06} and gives the
optimal protocols for the refrigeration cycle. The times spent on
the two isochores, $\tau_c$ and $\tau_h$, are not independent
variables as they satisfy the relation (\ref{gammac}).
 This is not surprising,  since the optimal protocols are fixed through
optimization on the $\chi$ function  as we have done.  When
$\gamma_c=\gamma_h$, the optimal times spent on the two isochores
satisfy the relation $\tau_c=\tau_h$.

\begin{figure}[htb]
\includegraphics[width=3.4in]{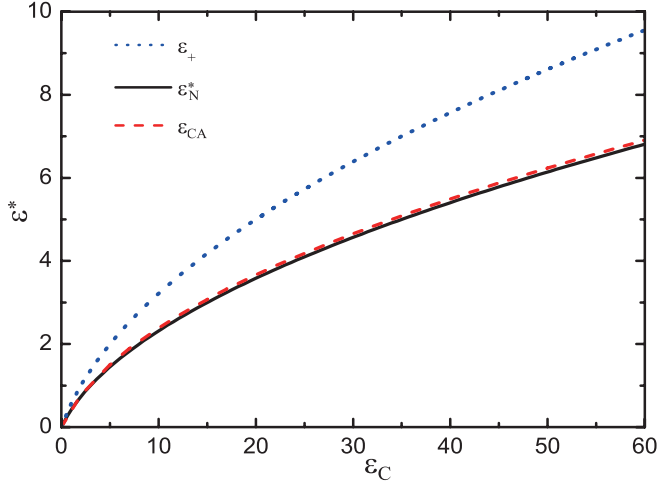}
 \caption{(Color online) COP at maximum $\chi$ figure of merit $\varepsilon^*$
   as a function of the Carnot COP $\varepsilon_C$.
 The numerical values of the COP, $\varepsilon^*_{N}$, are denoted by
 a black solid line and they are in nice agreement with the values of
$\varepsilon_{CA}$ which are denoted by red dashed line.   The
optimal COP obtained in the high-temperature limit, $\varepsilon^+$,
is represented by a blue dotted line. } \label{copf}
\end{figure}

Now consider the optimization on the external constrains of the
refrigerator in which the time taken for the adiabats $\tau_{adi}$
is assumed to be constant. Based on Eq. (\ref{chih}), optimizing the
figure of merit $\chi$ becomes equivalent to optimizing two bounds
of the energy unit $\omega_a$ and $\omega_b$. Extremal conditions
$\partial{\chi}/{\partial \omega_a}=0$ and $\partial{\chi}/{\partial
\omega_b}=0$ leads to the following relations:
\begin{equation}
\frac{{\beta _c x _c (\omega _b  - \omega _a )}}{{x _c  + 1}} =
\frac{{x _h  - x _c }}{{x _h  + 1}}\left( {\frac{{2\omega _b -
\omega _a }}{{\omega _a }}} \right),
\end{equation}
and
\begin{equation}
\frac{{\beta _h x _h (\omega _b  - \omega _a )}}{{x _h  + 1}} =
\frac{{x _h  - x _c }}{{x _c  + 1}}, \label{qiudaoxia}
\end{equation}
where we have used $x_c\equiv e^{-\beta_c\omega_a}$ and $x_h \equiv
e^{-\beta_h\omega_b}$. This set of two nonlinear equations can and
only can be solved numerically to yield the optimal values of
$\omega_a$ and $\omega_b$, provided that the temperatures of two
heat reservoirs $\beta_c$ and $\beta_h$ are given. In Fig.
\ref{copf} we plot the COP at maximum $\chi$ figure of merit,
$\varepsilon^*$, as a function of Carnot COP $\varepsilon_{C}$,
comparing  the CA COP $\varepsilon_{CA}=\sqrt{1+\varepsilon_C}-1$
with the values of $\varepsilon_+= ({\sqrt{9+8\varepsilon_C}}-3)/2$
(which is discussed below).  Figure \ref{copf} shows that our
numerical calculations ($\varepsilon^*_N$ ) are in nice agreement
with the values of $\varepsilon_{CA}$, which were  obtained
previously in low-dissipation Carnot-like refrigerators under
symmetric conditions \cite{Hu13, Tu12, Tom12} or endoreversible
refrigerators with Newton's heat transfer law \cite{Yan90}.

In the high-temperature limit when $\beta\omega\ll1$ and thus
$\tanh(\beta\omega/2)\simeq \beta\omega/2$, the heat transport law
is identified as the linear phenomenological law in irreversible
thermodynamics, since the amounts of heat exchanged during two
isochores, given by Eq. (\ref{qc}) and (\ref{qh}), simplify to
${Q_c}=\gamma_c\omega_a^2(\beta_c-\beta_h)f(\tau_c,\tau_h)/4$ and
$Q_h=\gamma_h\omega_b^2(\beta_c-\beta_h)f(\tau_c,\tau_h)/4$,
respectively. In such a case, substitution of the approximation of
$\tanh(\beta\omega/2)\simeq \beta\omega/2$ into Eq. (\ref{chih})
leads to $
\chi=\frac{f(\tau_c,\tau_h)\omega_a^{2}({\beta_c\omega_a}-{\beta_h\omega_b})}{{
{4(\omega_b-\omega_a)(\tau_{adi}+\tau_c+\tau_h)}}}.$ Although $\chi$
is a monotonically increasing function of $\omega_b$, one can
optimize $\chi$ in local region at given $\omega_b$ by setting
$\partial \chi/\partial \omega_a=0$, leading to the COP at maximum
$\chi$ figure of merit
\begin{equation}
\varepsilon^*=\varepsilon_+\equiv({\sqrt{9+8\varepsilon_C}}-3)/2.
\label{eps2}
\end{equation}
This result, reached in the high-temperature limit when the heat
transport law is linear phenomenological law in irreversible
thermodynamics, is particularly interesting. It is identical to a
reported universal upper bound that was derived in Refs. \cite{Tu12,
Hu13} using the low-dissipation assumption, and it also coincides
with the upper bound obtained in a minimally irreversible
refrigerator model \cite{Izu13}. In the high-temperature case when
in which the heat transport law is linear phenomenological law, our
refrigerator model reproduces the same upper bound  as one derived
from the low-dissipation refrigerators in the extremely asymmetric
limit. This  seems to imply that the same limit might be taken both
for the
 refrigerators with the linear phenomenological law and
for low-dissipation refrigerators in the asymmetric dissipation
limit. These values, however,  indicate greater validity to those
values obtained in previous papers \cite{Hu13, Tu12, Izu13}, as lies
in the fact that they were derived from the master equation
(\ref{ptpt}) based on stochastic processes.

The low-temperature limit when $\beta\gg1$ leads to
$\tanh(\beta\omega/2)$ approaches $1$. It is therefore indicated
that the amount of refrigeration per cycle, $Q_c$, becomes vanishing
and the refrigerator has lost its role in this case.

\section{COP at maximum $\chi$ derived from the refrigerator model under the
tight-coupling condition}

In order to study further the COP at maximum $\chi$ figure of merit,
we present the optimization on the performance of the cyclic
refrigerator model within the framework of irreversible
thermodynamics, identifying our model as a linear irreversible
refrigerator that works on the linear regime, close to thermal
equilibrium.

Since the working system  comes back to the original state after a
single cycle, the entropy production rate of the refrigeration cycle
is given by
$\dot{\sigma}={\beta_h}{\dot{Q}_h}-{\beta_c}{\dot{Q}_c}$, which
takes the form
\begin{equation}
\dot{\sigma}={\beta_c}{\dot{W}}+{\dot{Q}_h}\left({{\beta_h}
-{\beta_c}}\right),\label{sdt}
\end{equation}
where  the dot  ($\cdot$) represents the physical  quantity divided
by the cycle period $t_{cycle}$.  Within the context of irreversible
thermodynamics, the entropy production $\sigma$ can be expressed in
terms of the decomposition: $\dot{\sigma}=J_1 X_1+J_2 X_2$, where
$J_1 (J_2)$ denotes thermodynamic flux and $X_1 (X_2)$ is the
corresponding conjugate thermodynamic force. From this decomposition
and Eq. (\ref{sdt}),  we define  the thermodynamic fluxes
\cite{Izu10, Izu13, Sheng14}
\begin{equation}
J_1\equiv1/{t_{cycle}}, J_2\equiv\dot{Q}_h, \label{j1j2}
\end{equation}
and their conjugate thermodynamic forces
\begin{equation}
 X_1\equiv \beta_c W, X_2\equiv\beta_h-\beta_c. \label{x1x2}
\end{equation}

Theses  fluxes and forces can be described by using Onsager
relations as \cite{Bro05, On31}
\begin{equation}
J_1= L_{11}X_1+L_{12}X_2, \label{j1x2}
\end{equation}
\begin{equation}
J_2=L_{21}X_1+L_{22}X_2, \label{j2x2}
\end{equation}
where $L_{ij}$'s  are the Onsager coefficients with the symmetry
relation $L_{12} = L_{21}$ and they satisfy the constraints:
$L_{11}\geq 0, L_{22}\geq 0, L_{11}L_{22}- L_{12}L_{21}\geq 0$.
Noteworthy, linear irreversible thermodynamics described by the
linear relations between fluxes and forces [see Eqs. (\ref{j1x2})
and (\ref{j2x2})] holds well, provided that the systems are at local
equilibrium and they cannot operate at rates keeping them far away
from the limit of relatively near equilibrium \cite{Bro05, Rut09}.

 We
define $q\equiv L_{12}/\sqrt{L_{11}L_{22}}$ as the usual coupling
strength parameter and  find  $|q| \leq1$ from these constraints.
From Eqs. (\ref{j1x2}) and (\ref{j2x2}), we have
\begin{equation}
J_2=\frac{L_{21}}{L_{11}}J_1+(1-q^2)L_{22} X_2. \label{j2x}
\end{equation}

The heat current absorbed from the cold reservoir,
$\dot{Q}_c=\dot{Q}_h-\dot{W}$, can be expressed as
$\dot{Q}_c=J_2-X_1J_1/\beta_c$, and  the COP $\varepsilon$ for the
refrigerator is given by
\begin{equation}
\varepsilon=\frac{\dot{Q}_c}{\dot{W}}=\frac{\beta_c J_{2}}{X_1
J_{1}}-1, \label{copr}
\end{equation}
 where Eqs. (\ref{j1j2}) and
(\ref{x1x2}) have been used. It follows, substituting  Eq.
(\ref{j2x}) into Eq. (\ref{copr}), that the COP $\varepsilon$ takes
the form: $\varepsilon=\frac{\beta_c L_{12}}{X_1
L_{11}}+(1-q^2)\frac{L_{22} X_2}{J_1}-1$. In view of the fact that
$0<q^2<1$ and $X_2=\beta_h-\beta_c$ ($<0$), we find that the COP
$\varepsilon^*$ increases monotonically as the value of $q^2$
increases and approaches its maximum value when the tight-coupling
condition is satisfied with $|q|=1$. For the remainder of the paper,
we will discuss the COP at maximum $\chi$ figure of merit  for the
refrigerator which fulfills the tight-coupling condition $|q|=1$.

From Eq. (\ref{j2x}) and the tight-coupling condition $|q|=1$,  we
obtain the relation: $J_2=\lambda J_1$ with
$\lambda\equiv\frac{L_{12}}{L_{11}}$, a quantity independent of the
thermodynamic forces. For the tight-coupling refrigerator, the COP
and the target function, $\chi=\varepsilon \dot{Q}_c$, then become
\begin{equation}
\varepsilon=\frac{\beta_c \lambda}{X_1}-1, \label{copr2}
\end{equation}
\begin{equation}
\chi=\frac{\beta_c}{X_1}{\left(\lambda-\frac{X_1}{\beta_c}\right)}^2{J_1}
=\frac{\beta_c}{X_1}{{\left(\lambda-\frac{X_1}{\beta_c}\right)}^2}{L_{11}}
\left(X_1+{\lambda X_2}\right), \label{copr2}
\end{equation}
respectively. Setting $\frac{\partial{\chi}}{\partial X_1}=0$ for
given inverse temperatures $\beta_c$ and $\beta_h$, we obtain the
physical solution at $X_1=-(\sqrt{X_2^2-8\beta_c X_2}+X_2)\lambda/4
$. Substituting this solution into Eq. (\ref{copr2}), we find that
the COP at maximum $\chi$ figure of merit, $\varepsilon^*$, is also
given by Eq. (\ref{eps2}). Note that, this optimal value of
$\varepsilon$ is also the upper bound of the COP at maximum $\chi$
figure of merit, since the refrigerator model satisfy tight-coupling
condition $|q|=1$, which gives maximum  value of COP, as we
discussed below Eq. (\ref{copr}).

Before ending this section, we should emphasize that, as in previous
low-dissipation refrigerators \cite{Tu12, Hu13} and in minimally
nonlinear irreversible refrigerators \cite{Izu13} which were also
under the assumption of the local equilibrium, we re-derive the
upper bound $\varepsilon_+$ of optimal COP from the linear
irreversible refrigerators.  Whether under low-dissipation
assumption or within framework of irreversible thermodynamics,  the
phenomenological heat transfer laws are avoided for the cyclic
refrigerators.  However, the upper bound of the COP at maximum
$\chi$ figure of merit for refrigerators is the same as that derived
in the the refrigerators with a certain heat transfer law.  The
issue of exploring the intrinsic and universal relation between the
heat devices without use of heat transfer laws and those with
certain heat transfer laws may not be easy to address, but it
deserves to be studied in the future work. (Similar attempts have
been made to solve such a problem, and some interesting results have
been found for heat engines. See, for example, Ref. \cite{Wang12}).

\section{Conclusions}
In conclusion, we have established a quantum Otto refrigerator that
consists of two isochores (two constant frequencies $\omega_a$ and
$\omega_b$) and two adiabats by using a two-level atomic system as
the working substance. Employing finite-time thermodynamics, we
considered the COP at maximum $\chi$ figure of merit,
$\varepsilon^*$, for a quantum Otto refrigerator working with a
two-level atomic system, optimizing $\chi$ with respect two
frequencies $\omega_a$ and $\omega_b$. Our numerical calculations
show that the the values of $\varepsilon^*$ agree very well with the
CA values $\varepsilon_{CA}=\sqrt{1+ \varepsilon_{C}}-1$ at finite
temperatures. In the high-temperature limit, we obtained merely
considering $\omega_a$ as a freedom the COP at maximum $\chi$ as
$\varepsilon^*=\varepsilon_+=({\sqrt{9+8\varepsilon_C}}-3)/2$, which
is the upper bound of the optimal COP in low-dissipation or
minimally nonlinear irreversible refrigerators.  Within the
framework of irreversible thermodynamics, we showed that the COP at
maximum $\chi$ is also bounded from above the value of
$\varepsilon_+$, taking our model as a liner irreversible
refrigerator.

 \textbf{Acknowledgements}

This work is supported by the National Natural Science Foundation of
China under Grants No. 11265010, No. 11375045, and No. 11365015; the
State Key Programs of China under Grant No. 2012CB921604; and the
Jiangxi Provincial Natural Science Foundation under Grant No.
20132BAB212009, China.


\begin{thebibliography}{99}
\bibitem{notev} From another point of view, we find that the frequency
$\omega$ plays the role of the volume of the system (see in Ref.
\cite{8504} for more details). The frequency $\omega$ is, in fact,
the energy gap $\Delta \varepsilon$ between the ground and excited
states for the two state system. The frequency $\omega$ must be a
function of characteristic size (volume) $L$  of the system, and
contant $\omega$ means constant volume of the sytem. That is, the
isochoric process for the two level system is described by the
constant-frequency process.
\end{thebibliography}
\end{document}